# Experimental evidences of trions and Fermi edge singularity in single barrier GaAs/AlAs/GaAs heterostructure using photocapacitance spectroscopy

**Amit Bhunia[1], Mohit Kumar Singh[1], Y. Galvão Gobato [2], Mohamed Henini[3,4] and Shouvik Datta[1 *]**

[1]*Department of Physics &Center for Energy Science, Indian Institute of Science Education and Research, Pune 411008, Maharashtra, India*

[2]*Departamento de Física, Universidade Federal de São Carlos, 13560-905, São Carlos, SP, Brazil*

[3]*School of Physics and Astronomy, University of Nottingham, Nottingham NG7 2RD, UK*

[4]*UNESCO-UNISA Africa Chair in Nanosciences & Nanotechnology Laboratories, College of Graduate Studies, University of South Africa (UNISA), Muckleneuk Ridge, PO Box 392, Pretoria, South Africa*

***Corresponding author's email:** shouvik@iiserpune.ac.in

## Abstract

In this paper, we show how photocapacitance spectra can probe two dimensional excitonic complexes and Fermi edge singularity as a function of applied bias around 100 K. In lower density regimes ($<1\times10^{11}cm^{-2}$), the appearance of two distinct peaks in the spectra are identified as a signature of coexistence of both excitons and positively charged trions. We estimate the binding energy of these trions as ~2.0 meV. In the higher density regimes ($>1\times10^{11}$ $cm^{-2}$), we observe a sharp spectral transition from trions to asymmetric shaped Fermi edge singularity in the photocapacitance spectra around a particular reverse bias. However, these signatures are absent from the photoluminescence spectra measured under identical circumstances. Such dissimilarities clearly point out that different many body physics govern these two spectral measurements. We also argue why such quantum confined dipoles of spatially indirect trions can have thermodynamically finite probability to survive even around 100 K. Finally, our observations demonstrate that photocapacitance technique, which was seldom used to detect trions in the past, can also be useful to detect the traces of these spatially indirect excitonic complexes as well as Fermi edge singularity even at 100 K. This is mainly due to enhanced sensitivity of such dielectric measurements to dipolar changes within such heterojunction.

# I. INTRODUCTION

Experimental study of neutral excitons (say $X^0$) and charged excitons like positively or negatively charged trions ($X^+$ or $X^-$) inside two dimensional (2D) semiconductor quantum structures are becoming important for understanding the many body physics of excitonic complexes [1,2] as well as for novel applications of excitonic devices [3]. Existence of trions was first demonstrated in semiconductors by Lampert [4]. Since then, there are numerous reports on the spectroscopic signatures of trions at low temperatures of few Kelvins. This is mainly because of small binding energies (~1-2 mV) of trions in III-V material system [5-7]. Whereas, trions in 2D monolayers of transition metal di-chalcogenides (TMDC) can have much larger binding energies [8,9] ~20-50 meV. These large binding energies of excitons and trions in TMDC materials certainly made it much easier to study the many body physics of excitonic complexes even at room temperature. However, there are also reports of some disparities between theoretical and experimental observations of precise signature of positive or negative trions and their expected binding energies [7]. According to theoretical predictions [10], binding energy of positive trions should be higher due to larger hole effective mass but experiments had reported nearly similar values of binding energies [11] for both $X^+$ and $X^-$. At this point, we understand that identifying a trion as $X^+$ or $X^-$ depends mostly upon the device configurations and also on the dominant presence of either electrons or holes under external applied perturbation. The presence of these excitonic complexes usually occurs in moderate density regimes. Whereas many body interactions of charge carriers with Fermi sea lead to Fermi edge singularity (FES) in high density regimes. This was first predicted by Mahan [12] in degenerate semiconductors. Theory [13,14] and experiments [15-17] of such crossover to FES in different semiconductors were studied mostly by optical absorption and photoluminescence (PL).

We have already demonstrated [18] that photocapacitance can be used as a sensitive electro-optical tool to probe spatially indirect excitons formed across the AlAs potential barrier in GaAs/AlAs/GaAs heterostructure even at room temperature. In the present study, we report that photocapacitance spectroscopy can also be used to detect the specific signature of positively charged trions of indirect excitons ($IX^+$) at moderate levels of carrier densities ($<1\times10^{11}$ $cm^{-2}$) as well as Fermi edge singularities (FESs) in higher level of carrier densities ($>1\times10^{11}$ $cm^{-2}$) in a single barrier GaAs/AlAs p-i-n heterostructure. Later, we argue why photocapacitance can sense these $IX^+$s even at 100 K, which was not possible to detect with photoluminescence. Formations of triangular quantum wells (TQWs) around Γ-AlAs potential barrier also help these spatially indirect trions to survive even at this moderately low temperature. Therefore, we demonstrate that photocapacitance is a more sensitive experimental technique for detecting trions and many body effects like FES than PL in such heterostructures.

## II. SAMPLE AND EXPERIMENTAL METHODS

The heterostructure used in this work was grown by molecular beam epitaxy on a semi-insulating GaAs (311)A substrate. A highly doped 1.5 μm buffer layer of p-GaAs ($4\times10^{18}$ $cm^{-3}$) was first grown on the substrate. This layer was used as the bottom electrical contact for our measurements. This was followed by a 100 nm thick p-GaAs ($1\times10^{17}$ $cm^{-3}$) layer. Then an 8 nm thick, undoped AlAs potential barrier having 100 nm thick undoped GaAs spacer layers on both sides were grown. Finally, a 100 nm n-GaAs ($2\times10^{16}$ $cm^{-3}$) layer and a 0.5 μm highly n-doped GaAs ($4\times10^{18}$ $cm^{-3}$) capping layer were grown to complete this heterostructure. Arrays of circular gold mesas with 400μm diameter and area of ~$5x10^{-4}$ $cm^2$ were fabricated as top metal contacts. These also facilitated optical access from the top of the device.

We used Agilent's E4980A LCR meter with 30 mV of rms voltage at 200 Hz frequency (unless mentioned otherwise) for photocapacitance measurements. Photocurrent measurements were performed using the same LCR meter in the DC mode as well as with Keithley 2611 source meter. An Acton Research's SP2555 monochromator having 0.5 m focal length (with $\Delta\lambda$ ~ 1.5 nm) along with 1000 Watts quartz-tungsten-halogen lamp was used for spectroscopy. Spectral response of lamp, monochromator combination was reasonably smooth and changes slowly and monotonically within the wavelength ranges used in our experiments. PL spectra were measured using He-Ne gas laser and CCS200 compact spectrometer from Thorlabs with spectral accuracy <2 nm FWHM @633 nm. For temperature variation, we used a closed cycle Helium gas CS-204S-DMX-20 cryostat from Advance Research Systems along with the Lakeshore (Model-340) temperature controller.

We illuminated our sample from the top n-GaAs side and simultaneously applied reverse bias. Photocapacitance spectra at 100 K were measured under an optical intensity of 17.4 $\mu W/cm^2$ at peak wavelength of 830 nm under reverse bias. In this work, we will only focus on the sharp resonant peak like spectral features in photocapacitance under non-zero biases. We carefully selected top mesa contacts for which DC-photocurrents are restricted within nano-Ampere ranges under reasonably small applied biases of a few volts. This prevents unwarranted dielectric screening of Coulomb attractions required to form these excitonic complexes. The applied electric fields corresponding to each bias have been estimated by dividing the bias magnitude with the thickness (~208 nm) of the intrinsic region of the heterostructure.

## III. RESULTS AND DISCUSSIONS

### A. Bias dependent splitting in photocapacitance spectra leads to positive trions and Fermi edge singularity

In the present article, we investigated bias induced formation dynamics of spatially indirect excitons and trions near the GaAs/AlAs heterojunction under light and also under applied reverse biases around 100K. In Fig. 1(a), changes in photocapacitance spectra under different applied reverse biases are shown. As the magnitude of reverse bias increases from -0.1 V to higher values, single peak like excitonic spectra is split up into two distinctly separate peaks. This kind of peak splitting with increasing bias was clearly absent at room temperature [18]. Initially the spectral shapes of these peaks remain nearly symmetric until the bias value of -0.6 V. After this, an asymmetric spectral shape emerges where the low energy spectral tails rise slowly and the high energy side falls sharply. In addition, an enhancement of the low energy peak is clearly noticeable around -0.8 V and beyond. On the other hand, the higher energy peak almost vanishes gradually at larger biases.

To understand the above observations, a schematic band diagram of the GaAs/AlAs/GaAs heterojunction is drawn in Fig. 1(b) following previous reports [18,19,20]. We will not go into details of these X and Γ valley contributions on the carrier transport which was already well investigated and discussed earlier[18]. Under the above mentioned experimental conditions, we expect [18] the formation of hole accumulation layers in the form of 2D hole gas (2DHG)within the GaAs TQW near Γ-AlAs barrier in the top n-type GaAs layer. Similarly, we expect to have electron accumulation layer as 2D electron gas (2DEG) forming inside the TQW on the p-type GaAs side. These accumulated electrons and holes can form hydrogenic bound states of indirect excitons ($IX^0$) due to their mutual Coulomb attractions as reported earlier [18]. This schematic diagram matches with the standard solution of self-consistent one dimensional Schrodinger and Poisson equations as described in earlier reports [21,22]. The 2DEG is spatially separated by 8 nm thin AlAs barrier from the 2DHG. The presence of this extra potential barrier in tunnel diode structure may introduce further complexities. Additional charge accumulations in TQWs can lead to solutions of Poisson equation which are not [21] fully self-consistent to the Schrodinger equation for such resonant tunneling structures.

We understand that large numbers of electrons and holes accumulate in these 2DEG and 2DHG, respectively, with increasing reverse biases. As a result, corresponding carrier densities in each TQW also increase. The charge density of photo generated carriers per unit area ($\sigma_{ph}$) around the heterojunction for each bias values are estimated from Fig. 1(a) using the following relationship [8]

$$\sigma_{ph} e = CV \tag{1}$$

where $e$ is the electric charge and $C$ is the peak value of excitonic photocapacitance per unit area at each applied bias ($V$). Following the description in Fig. 1(b), it is important to note here that one can also relate this $\sigma_{ph}$ to the density of dipolar charges per unit area in a parallel plate capacitor configuration as

$$\sigma_{Ph} = \vec{P}.\hat{z} \tag{2}$$

where $\vec{P}$ is the effective polarization of these excitonic dipoles and $\hat{z}$ is the unit vector along the growth direction. Due to the presence of Γ-AlAs potential barrier and both types of carrier accumulation in these TQWs, we purposely avoided any quantitative estimation of carrier density using standard capacitance-voltage plot which is strictly based on depletion approximation. However, at high enough injected carrier density (>$3\times10^{10}$ $cm^{-2}$ or reverse biases >|-0.3| V as shown in Fig. 1(a)), another peak at the low energy side starts to appear. Evidently, concentrations of holes in the 2DHG become larger compared to those of electrons in 2DEG. This is because the sample is optically excited from the top n-GaAs side only and bias is also applied in the reverse direction. Such excess of one type of carriers (in our case holes) favor the formation of positively charged trions which can be light or voltage controlled [23]. As a result of this excess hole density in 2DHG, we now ascribe the sharp resonant spectral features on the low energy side of the spectra with positively charged indirect excitons or trions ($IX^+$). Until the bias value reaches -0.6 V or carrier density ~$8\times10^{10}$ $cm^{-2}$, both peaks look nearly symmetric. Even though, we see that the trion peak is getting bigger than the exciton peak. This low energy $IX^+$ state is more energetically favorable and tends to dominate at higher applied biases. Subsequently, after the reverse bias around -0.8 V, we observe a drastic change in the spectral behavior as evident from Fig. 1(a). The low energy trion peak indeed rises significantly whereas the exciton peak almost vanishes around the density value of $1.2\times10^{11}$ $cm^{-2}$. Moreover, the spectral shape of this trion peak gradually becomes asymmetric in nature. This type of spectral transition is usually [15-17,24] identified with Fermi edge singularities at high enough carrier densities. Correspondingly, in our photocapacitance spectra, we see similar sharp, asymmetric enhancement of the trion peak once the photo generated carrier density approaches ~$1\times10^{11}$ $cm^{-2}$. In the present case, it is likely that at high photo generated carrier densities, localized hole states of 2DHG or holes in shallow impurity states can experience enhanced Coulomb attraction with the Fermi sea of electrons in 2DEG at such low temperatures. As a result, we observe a significant enhancement of the absorption edge of this trion related FES. Previously, Yusa et al. [15] had reported asymmetric line shape in the absorption spectra which showed fast rise in the low energy side and slow fall in the high energy side. Chen et al.[16] and Skolnick et al. [17] had observed slow rise in the low energy side and fast fall in the high energy side of the PL spectra. Usually in the high density limit, enhanced carrier-carrier scattering broaden the high energy tail of FES. However, we notice that photocapacitance spectra in Fig. 1(a) have much sharper high energy edge in a system consisting of both 2DEG and 2DHG. In addition, we still see the remnant excitonic state above the FES. Although this excitonic state is almost vanishing but due to the continued presence of this state, the high energy side of the main FES peak may possibly remain sharp too. We speculate that sub-band gap localized states may significantly modify the photocapacitance spectral broadening in the lower energy side as shown in Fig. 1(c).

Nevertheless, in the common understanding of FES inside a quantum well structure, usually a sea of electrons in the conduction band is attracted to a single localized hole state near the valence band. This usually shows a power law singularity [15] of the form

$$A(\omega) = \frac{1}{|\omega - \omega_0|^{\alpha}} \tag{3}$$

in the absorption spectra where $\omega$ is the angular frequency of incident light, $\omega_0$ is the resonant angular frequency of the absorption peak and $\alpha$ is a power law exponent. At -0.8 V, we found $\alpha$=~0.4, albeit from the low energy tail of the FES spectra. This value, however, matches with the literature [15]. At this stage, it is important to reiterate that here we are actually dealing with the dual presence of both 2DEG and 2DHG. In a way, we are looking at how a Fermi sea of electrons inside a 2DEG interacts with excess holes localized around 2DHG. We will explain this in more detail at the end of this section.

We also note that the full widths at half maxima (FWHM ~8 meV) of these FES peaks are somewhat lower than the corresponding thermal width ~$k_BT$=8.7 meV at 100K. We can assume that this FWHM may provide an estimate of the Fermi Energy following the discussion by Skolnick et al. [17]. The 2D Fermi energy is usually calculated using the expression of (for spin ½ particles)

$$E_F = \frac{\pi \hbar^2 \, \sigma_{ph}}{m^*} \tag{4}$$

where ħ is the reduced Planck's constant, $\sigma_{ph}$ is the dipolar charge density per $cm^2$. Here, for these excitonic dipoles around the AlAs barrier, we have used $m^* = \mu$ as the reduced mass of holes and electrons ~$0.056 m_0$ in GaAs, $m_0$ is the free electron mass. Therefore, using Eq. (4), we find that FWHM $\sim E_F = 8\, meV$ corresponds to a carrier density of ~$1.86 \times 10^{11}$ $cm^{-2}$ (following the discussion of Skolnick et al. in Ref. [17]). This estimate matches closely with ~$1.82 \times 10^{11}$ $cm^{-2}$,deduced from Eq. (1) at a bias of -1.0 V around where the prominent signatures of FES like asymmetric spectra is visible.

In TQWs, we understand that only the first few states of these 2DEG or 2DHG remain strongly quantum confined, i.e. behave as two dimensional in nature. Higher energy states in TQWs are less localized as the well width becomes wider [21]. We understand this from the quantized energy $E_n$ and nth eigenfunctions $\psi_n(z)$ of an infinite barrier TQW [22], which are given by

$$E_n \sim \sqrt[3]{\left[\left(\frac{\hbar^2}{2m^*}\right)\left[\frac{3\pi e F_z}{2}\left(n - \frac{1}{4}\right)\right]^2\right]} \tag{5}$$

and

$$\psi_n(z) = Ai\left[\frac{2m^* e F_z}{\hbar^2}\left(z - \frac{E_n}{eF_z}\right)\right] \tag{6}$$

where n = 1,2,3…, Ai(z) is the Airy functions, $m^*$ is the effective mass of electron/hole, e is the electronic charge, $F_z$ is the electric field along the growth direction along $z$ axis, $\hbar$ is the reduced Planck's constant. The ratio of the first excited energy to the ground state energy is $\frac{E_2}{E_1} = \sqrt[2/3]{\left(\frac{7}{3}\right)} \sim 1.76$. Therefore, it is expected that in practice, with a finite barrier TQW like those shown in Fig. 1(b), only the ground state energy may remain bound and quantum confined. However, with increasing bias and charge accumulations around the AlAs barrier, the quasi Fermi levels can suddenly penetrate these TQWs at the onset of FES. As a result, we also observe a sharp transition around a reverse bias of -0.8V where the exciton-trion spectra suddenly convert into asymmetric FES spectra. This is because, with increasing bias, the quasi Fermi level $E_{Fn}$ ($E_{Fp}$) of electrons (holes) can get pinned with the discrete ground state energy of 2DEG(2DHG) in these TQWs. However, the quasi Fermi Level separation $\left(E_{Fn} - E_{Fp}\right)$ can vary along with the ground state energy levels of these electron and hole TQWs as shown in Fig. 1(c). As a result, we find that $\left(E_{Fn} - E_{Fp}\right)$ can keep on increasing with the increase of reverse bias (electric field) as per Eq. (5). In the absence of any quantum confined excited states in these shallow TQWs, the maximum optical absorption energy for indirect excitons can then be limited by

$$E_{Opt}^{Max} = E_g + E_1^e + E_1^{hh} \equiv \left(E_{Fn} - E_{Fp}\right) \quad (7)$$

where $E_g$ is the bulk bandgap of GaAs, $E_1^e (E_1^{hh})$ is ground state energy of electrons (heavy holes). In fact, we observe FES related enhancement at this $E_{Opt}^{Max}$ edge only. In Fig. 1(c), we pictorially depict this FES at $E_{Opt}^{Max}$ following our schematic diagram given in Fig. 1(b). Moreover, we assume that quantum confined holes in 2DHG can be further localized to shallow impurity states or interface trap states or to states related to alloy disorder inside 2DHG near the valence band edge. Any optical absorption from such shallow states near the 2DHG to $E_1^e$ can couple the holes left in these localized states with the 2DEG to create the FES. This can also explain the broad low energy tail of the FES spectra (Fig. 1(a)). This is also schematically described in Fig. 1(c) and fully support our above explanations based on the formation of positively charged $IX^+$ at lower bias levels.

### B. Red shift and binding energy of indirect trions at 100K

To picture the above understandings, in Fig. 2(a), we use a 2D color filled contour plot of Fig. 1(a). This graphic presentation is essential for better visibility of spectral splitting, red shift of low energy $IX^+$ peak and energy separation of both exciton and trion peaks. Eventually the low energy trion peak leads to FES with increasing reverse biases. The dashed lines are drawn to guide the eyes only. A clear red shift of $IX^+$ spectra with the applied bias is visible as mentioned above, whereas there is hardly any shift in peak energy for $IX^0$ line. At this stage, it is not clear why excitonic peak at slightly higher energy is not at all affected by increasing the electric field. Besides, applied voltage dependent spectral energy separation between two peaks clearly changes and increases with increasing bias as evidenced from the two dashed lines drawn in Fig. 2(a).

Following Bugajski and Reginski [25], we now discuss the transformation of trions into FES spectra. In the parabolic band approximation, the Fermi energy $E_{Fn}$ of electrons in the bulk can be written in the form of low energy onset of absorption ($\varepsilon_F$) and emission edge ($\varepsilon_g$) as

$$E_{Fn} = (\varepsilon_F - \varepsilon_g)\left(1 + \frac{m_e^*}{m_h^*}\right)^{-1} \quad (8)$$

where $\left(\frac{m_e^*}{m_h^*}\right)$ is the effective mass ratio of electron and hole of GaAs. The Fermi energy varies with photo generated carrier density. As discussed above, we are driving both photo generated electrons and holes towards the AlAs barrier. So, the gap between the quasi Fermi levels $(E_{Fn} - E_{Fp})$should increase with increasing bias. However, this change in $E_{Fn}$ can be more in comparison to changes in $E_{Fp}$, because, in general, Fermi levels are inversely proportional to effective mass of the respective carrier and $\left(\frac{m_e^*}{m_h^*}\right) \ll 1$. As there is no observed change in the low energy onset $\varepsilon_g$ of photoluminescence (as will be shown later) with increasing reverse bias, we may infer that $\varepsilon_F$ may actually blue shift with increasing bias. Here we assume that any change in the effective mass ratio as a result of bias induced changes in the shape and size of TQWs is negligible. Consequently, FES related enhancement of photocapacitance spectra at $E_{Opt}^{Max} = (E_{Fn} - E_{Fp})$ should also show blue shift with increasing reverse bias greater than -0.8 V. However, we see the opposite effect and the high energy edge of FES actually red shifts. As mentioned above, in this case FES is a many body excitonic effect involving both 2DEG and 2DHG. Therefore, understanding the red shift of FES peak will not be possible without considering how these indirect excitonic complexes and Fermi levels are evolving with increasing bias in such heterostructure. From our recent report [18] on room temperature indirect excitons, we understand that the electrons and holes of $IX^+$ can come closer with increasing bias. Consequently the binding energy of this $IX^+$ can increase with increasing bias. This can lead to the observed red shift of this trion FES edge. On the other hand, this significant ~10 meV red shift of $IX^+$ line with the applied bias in this coupled TQWs structure may be related to quantum confined Stark effect (QCSE) [20, 21] where exciton peaks usually broaden with increasing electric field. This can also be due to a bias induced reduction of electron-hole separation across the AlAs barrier as reported [18] in our recent work at room temperature.

The magnitude of this peak splitting initially increases slowly and then in the higher bias regime it varies significantly as also reported by Yusa et.al. [15]. With increasing bias the carrier density also increases which eventually contributes to strong scattering mechanisms [21]. However, rigorous explanation of spectral broadening of FES requires understanding the full picture of many body scattering mechanisms and extrinsic carrier dynamics within such heterostructure, which is currently beyond the scope of this study. In Fig. 2(b), spectral red shift of the trion peak energy has been fitted with linear dipole moment equation

$$E = E_0 + \vec{p}.\vec{F} \tag{9}$$

where E is peak energy of $IX^+$, $E_0$ is the zero field energy, $\vec{p}$ is the effective permanent dipole moment of these trions and $\vec{F}$ is the applied electric field. We can see that $\vec{p}$ and $\vec{F}$ are directed opposite [18] to each other (see Fig. 1(b)). In addition, $\vec{p}$ is anticipated to be independent of $\vec{F}$ at such low temperature because of the observed linear variation of E with $\vec{F}$ (see Fig. 2(b)). Two regions of the bias dependent photocapacitance spectra have been fitted with two separate straight lines with slopes $m_1$ and $m_2$. These slopes actually estimate the values of respective dipole moments. Slope $m_1$ corresponds to the dipole moment value ~$2.96\times10^{-28}$ C-m before the onset of FES. However, we get substantially larger value of this dipole moment from slope $m_2$ ~ $7.97\times10^{-28}$ C-m for higher bias region when FES get started. In higher electric field regime, there will be more carrier injection to the TQWs, which eventually contributes to larger dipole moment compare to the lower bias one.

As can be seen in Fig. 2(c), the peak intensity ratio of the peaks $IX^+$ and $IX^0$ varies slowly at the beginning. It then changes significantly with increasing applied electric fields. The ratio remains near unity in the low bias side, indicating equal contribution from both excitons and trions. However, with increasing biases, the trions start to dominate the photocapacitance spectra. This is because positive trions formation is always more energetically favorable compared to the excitons. As a result, the peak photocapacitance ratio of trions and excitons is shown to vary with the applied field, implying the formation of trions at the expense of excitons with the increase of electric field in the reverse bias regime. We tried to fit the above variation with a modified Arrhenius rate ($R^*$) equation [26]

$$R^* \approx \frac{1}{\tau} = \nu^* \exp\left( -\frac{E_a}{k_B} \eta V_{dc} \right) \tag{10}$$

where $\eta$ is a proportionality factor in units of $V^{-1}K^{-1}$ to ensure correct dimensionality and $E_a$ is activation energy for transitions activated by bias voltage $V_{dc}$, $\nu^*$ is the thermal pre-factor representing the heat bath, $\tau$ is the time period of the transition and $k_B$ is the Boltzmann constant. The experimental data on peak photocapacitance ratio in Fig. 2(c) fits well with such bias activated model. The ratio drastically increases around the reverse bias of -0.8 V.

We observe that exciton peaks gradually decrease and hardly shift in energy with increasing applied bias, whereas the trion peaks slowly enhance and red shift drastically after certain applied bias. This can happen if more and more positively charged trions ($IX^+$) form at the expense of excitons ($IX^0$) in the presence of excess holes. Moreover, the trion peak ultimately converts to FES after a certain density regime which is in our case around $1\times10^{11}$ cm$^{-2}$. This understanding matches well with explanations provided by Huard et.al. [27]. Accordingly, we plot the energy difference between two peaks with the estimated Fermi energy for each applied bias in Fig. 3. This is then linearly fitted with the following relation [8,27].

$$E_{IX^0} - E_{IX^+} = E^b{}_{IX^+} + E_{Fp} \tag{11}$$

Where $E^b{}_{IX^+}$ is the positive trion binding energy, $E_{IX^0}$ and $E_{IX^+}$ are peak energies for indirect excitons and positive trions and $E_{Fp}$ is the Fermi energy change due to extra holes in the valence band. $E_{Fp}$ is estimated from photo generated carrier density using Eq. (4) with effective mass of hole as $m^*_h \sim 0.45\ m_0$ in GaAs. From the intercept, we estimate the value of an average trion binding energy of ~2.0 meV. We are also assuming that the estimated value of $E^b{}_{IX^+}$ using the above Eq. (11) is independent of applied bias. Interestingly, it matches well with earlier reported value in GaAs/AlAs system [5-7,11]. This supports our spectroscopic confirmation of the presence of trions ($IX^+$) using photocapacitance. Whereas, mostly reports estimate this binding energy from PL measurements, here we show that trion binding energy can also be determined from photocapacitance data also.

Moreover, dissociation of either excitons or trions inside a solid is always a many body statistical process with thermodynamic dissociation probability $\sim \exp\left(-\frac{E^b}{k_B T}\right)$, where $E^b$ can be either the exciton or the trion binding energy [23,28,29]. This is because [27, 28], such dissociations are governed by the following Saha's ionization equations.

$$\frac{n_h n_e}{n_{X^0}} = \frac{C_{X^0}^2}{C_e^2 C_h^2}\ exp\ \left(-\frac{E^b{}_X}{k_B T}\right) \tag{12}$$

$$\frac{n_X n_h}{n_{X^+}} = \frac{C_{X^+}^2}{C_X^2 C_h^2}\ exp\ \left(-\frac{E^b{}_{X^+}}{k_B T}\right) \tag{13}$$

where $E^b{}_{X^0}$, $E^b{}_{X^+}$ are binding energies of excitons and positively charged trions, $n_i$ are carrier densities, $C_i = \hbar\sqrt{\frac{2\pi}{m_i k_B T}}$ are thermal wavelengths and $m_i$ are effective masses of (i = e, h, $X^0$, $X^+$) electrons, holes, excitons and trions, respectively. This clearly shows that even when the binding energy $E^b{}_i$ is less than the thermal fluctuation energy ($k_BT$), there still can be a finite, non-zero probability of having some bound states of excitons or trions. It is remarkable to discover that photocapacitance spectroscopy is still sensitive to these very small numbers of excitons or trions.

### C. Differences in spectra at a fixed bias and with respect to temperature

Photocapacitance and photo-G/ω spectra at 200 Hz are compared with DC-photocurrent spectra for -0.6 V bias at 100K as shown in Fig. 4(a). Spectral splitting is only observed in photocapacitance spectra whereas we don't see any spectral splitting in photocurrent and photo-G/ω. Also there is a slight shift in peak energy position observed among these. Such differences were also observed at room temperature and were attributed to dissociation of excitons within the heterojunction but only from those direct excitons which are not closer to the AlAs potential barrier as explained before [18]. We would like to mention here that a simple equivalent circuit consisting of capacitance (C) and conductance (G) in parallel was used to extract photocapacitance using LCR meter such that |C| is always greater than |0.1×G/ω| where $\omega=2\pi f$. We know [18] that photocapacitance arises mainly from the bias driven accumulated charge carriers near the AlAs barrier whereas the DC-photocurrent is solely affected by the charge carriers driven out of the junction. However, contributions to photo-G/ω can come from both these regions. Moreover, one can notice that the spectral broadening in the photocurrent spectra is reasonably higher than the photo-G/ω and photocapacitance at -0.6 V. This certainly implies that the origin of DC-photocurrent spectra is different. The measured photocurrent is likely to originate from the dissociation of excitonic [30] photo absorptions which are mostly forming away from the junction [18]. This will be further elaborated in Sec. III D.

In Fig. 4(b), we now explain how the temperature variation changes the photocapacitance spectral shape and the peak energy for this particular applied bias of -0.6 V. Here, we observe that the single excitonic peak at room temperature drastically changes to narrower double peaks of both excitons and trions at 100 K. Frequency and temperature always play complementary roles in the measurement of capacitance. Comparisons with bias dependent photocapacitance spectra are shown in Fig. 1(a). This clearly indicates that such peak splitting occur only at low temperature and also under high applied biases. Such a critical role of applied biases will again be emphasized in the next Sec. III D. Blue shift of the spectral peak energies with decreasing temperature can be explained from the usual band gap expansion.

### D. Advantages of photocapacitance spectra in comparison to PL spectra in studying such indirect excitonic complexes.

In most studies, PL spectra were mainly used to understand the intricate details of many body carrier dynamics of trions and FES. In Fig. 5(a) we plot the PL spectra at 100 K for different applied biases. The spectral broadening, i.e. full width at half maxima (FWHM), is certainly more than that of usual intrinsic FWHM of any excitonic PL at low temperature. However, unlike photocapacitance spectra, we do not observe spectral splitting and any spectral shift in the PL spectra under increasing reverse biases. Also the spectral peak energy is expectedly at a lower energy in comparison to the photocapacitance spectral peak under similar experimental conditions. This definitely indicates two different physical origins for PL and photocapacitance spectra. Integrated PL intensity versus optical excitation intensity also follows a linear variation in log-log scale which gives a slope of ~ 0.96 at 100 K indicating [26] excitons as shown in Fig. 5(b). PL spectra are mostly unaffected by bias. PL is also a two-step probabilistic process involving photoexcitation and radiative recombination. It is expected that neutral excitons ($IX^0$s) and positively charged trions ($IX^+$s) around the AlAs barrier hardly contribute to PL due to the lack of spatial overlap of their respective electron and hole wave functions. Therefore, the observed PL is mainly due to the presence of excitons which are photo excited far away from the AlAs junction and then radiatively recombine. As a result, PL spectra are hardly affected by the applied bias whereas the indirect trions ($IX^+$) forming near the junction are strongly affected by the applied bias while their dipolar contributions are being sensed by photocapacitance. At this stage, we strongly suggest that photocapacitance is a much better suited and elegant technique to probe the experimental signature of these indirect excitonic complexes rather than the conventional PL.

In addition, optical intensity dependence of photocapacitance spectra has been studied for fixed biases at 100 K. Spectral peak splitting and shapes are observed to be hardly affected by the variation of the optical excitation intensity over the range 2 to 50 μW/cm$^2$.Increase of the optical excitation intensity only increases the magnitude of the overall photocapacitance. Interestingly, it is the applied bias which plays a more crucial role to inject the optically generated carriers to respective TQWs to form excitonic complexes. Therefore, we now have experimental proof that formations of trions and FES are predominantly controlled by the applied reverse bias and not so much by the optical excitation intensity.

## IV. CONCLUSIONS.

In summary, photocapacitance was used to detect the formation of spatially indirect, bias driven, positively charged trions and excitons around AlAs barrier in GaAs/AlAs single barrier samples even at 100 K. Contrary to popular beliefs, the presence of trions at such a high temperature of ~100 K is explained using the Saha's ionization equation. This indicates that there will always be a small but finite thermodynamic probability of existence of excitonic complexes having binding energy lower than the thermal bath energy ($k_BT$). Bias dependent spectral peak splitting is also demonstrated. Moreover, in the higher density limit $\geq 1\times10^{11}$ $cm^{-2}$, we have also identified the signatures of Fermi edge singularity from photocapacitance spectra even at 100 K. Interestingly, we have shown that these small fractions of trions and excitons do respond to capacitance measurements but not to conventional optical measurements like PL and DC-photocurrent. Although, researchers mostly use PL to study these excitons and trions in a variety of materials it is well established here that photocapacitance is a superior and sensitive experimental tool to probe and study these types of spatially indirect excitonic complexes in these single barrier p-i-n heterostructures. Moreover, the existence of these delicate excitonic complexes at such elevated temperatures will pave the way for their direct experimental manipulations leading to novel applications in next generation optoelectronics and telecommunications.

**ACKNOWLEDGRMENTS**

Authors acknowledge Department of Science and Technology, India (Research Grants # SR/S2/CMP-72/2012 and SR/NM/TP13/2016). We are grateful to Prof. B. M. Arora from IIT-Mumbai for his advice on band diagrams. AB and MKS are thankful to DST, India for Inspire PhD Fellowship and IISER-Pune for Integrated PhD Fellowship, respectively. YGG acknowledges the financial support from the Brazilian agency Fundação de Amparo a Pesquisa do Estado de São Paulo (FAPESP) (Research Grant # 16/10668-7). MH acknowledges support from the UK Engineering and Physical Sciences Research Council.

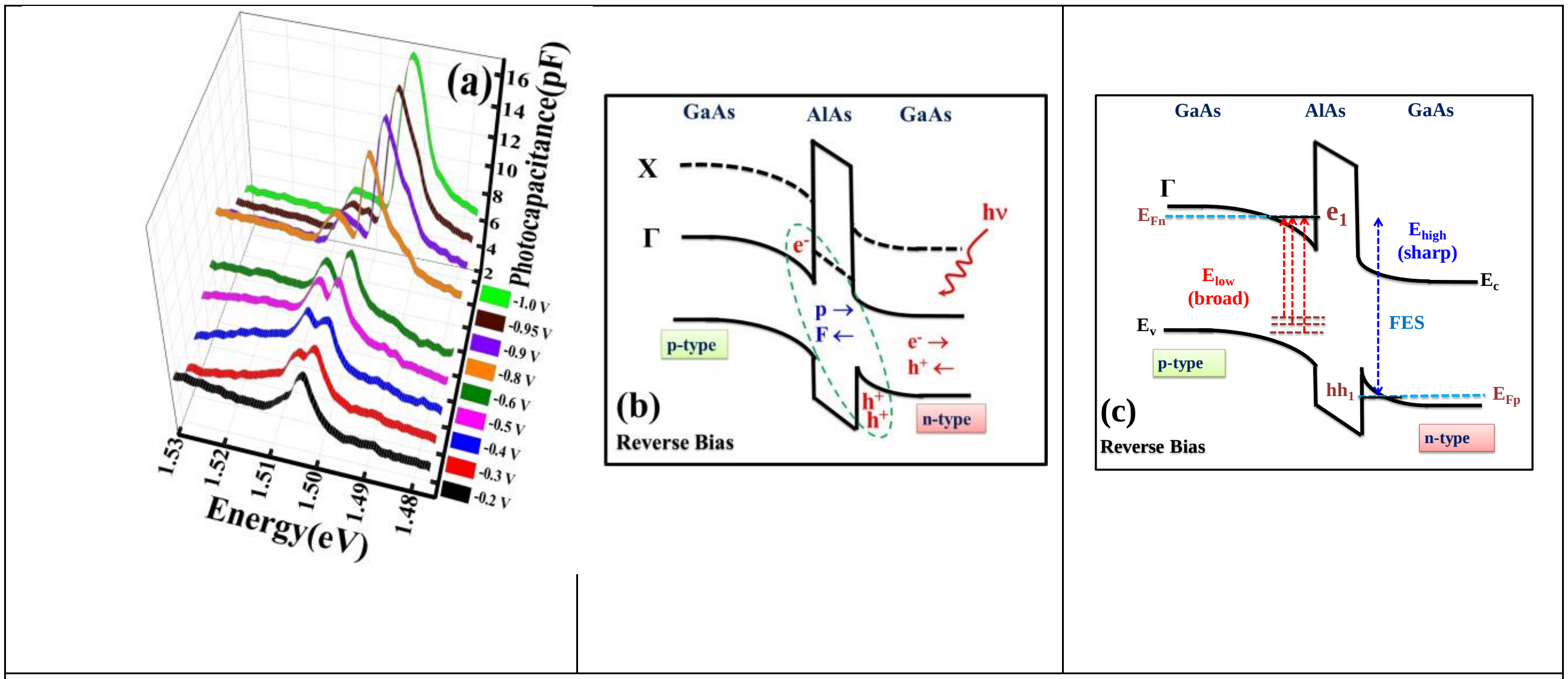

FIG. 1. (a) Variation of excitonic photocapacitance spectra with optical excitation energy for different reverse biases at 100 K. Single excitonic peak splits up into two distinct peaks of exciton and trion with increasing applied biases. Low energy trion peak evolves into FES like asymmetric spectra at higher biases. The last four spectra from -0.8 V to -1.0 V are vertically shifted by 2 pF for better visibility. (b) Schematic band diagram of the sample under light illumination and reverse bias condition to explain the formation of positively charged trions. (c) Origin of low and high energy tails of the FES spectra is schematically explained using a similar diagram. Relevant transitions contributing to low and high energy optical absorption tails are depicted respectively with red and blue arrows.

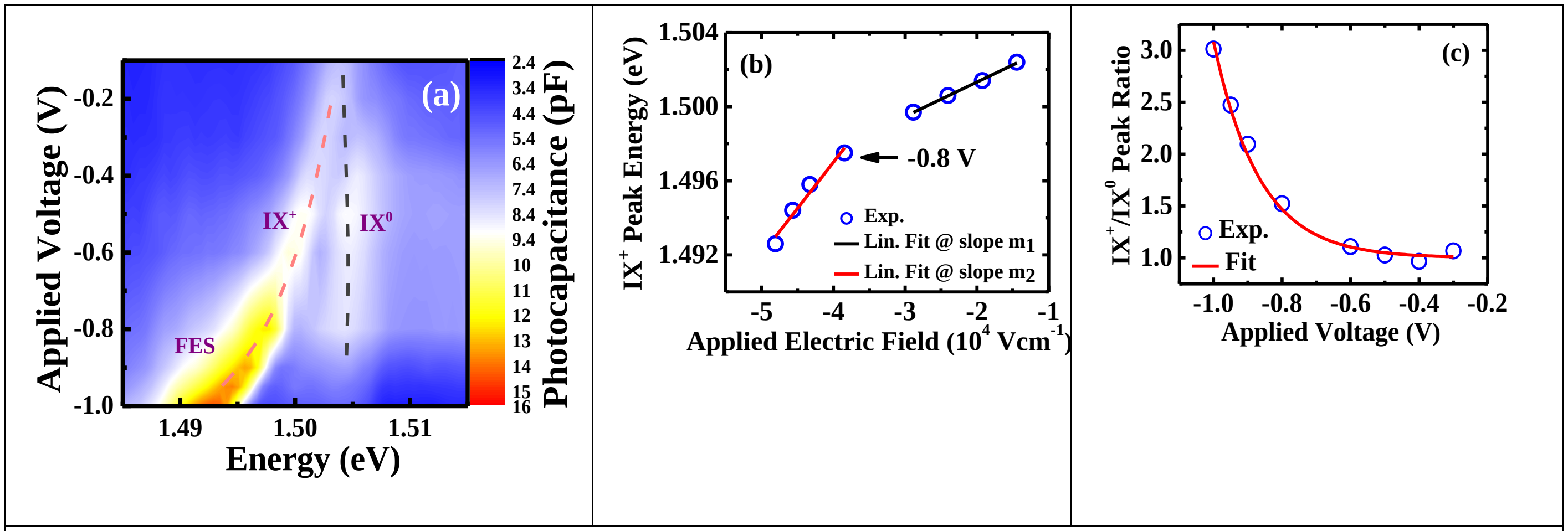

FIG. 2. (a) 2D colored contour plot of Fig. 1(a) is drawn. Two dashed lines are guiding to eyes only. Significant red shift of the $IX^+$ line and the visible splitting of the spectra with bias are clearly pointed out. In higher bias, the symmetric $IX^+$ line turns into Fermi edge singularity (FES) as the spectral shape become asymmetric. (b) Linear variation of peak energy of positively charged trion with increasing electric field is demonstrated to estimate the permanent dipole moments. There is a clear discontinuity around -0.8 V with a field of $-3.8\times10^4$ V.cm$^{-1}$. This represents a distinct transition from trion like peak to FES like spectral signature. (c) Bias activated changes in exciton and trion component of the spectra is shown.

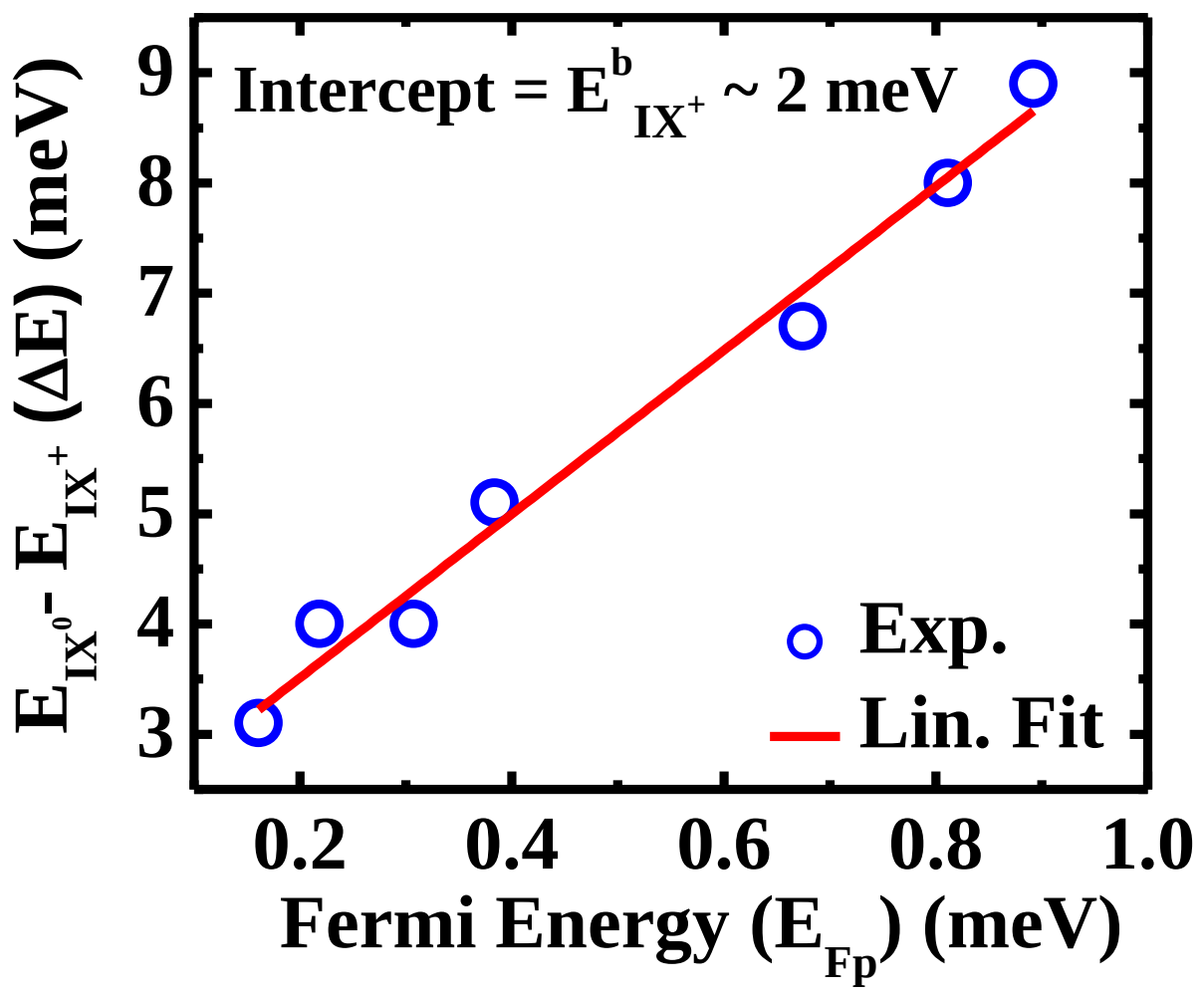

FIG. 3. Variation of peak splitting energy for excitons and trions with respect to the Fermi energy. The estimated trion binding energy matches well with past results in GaAs.

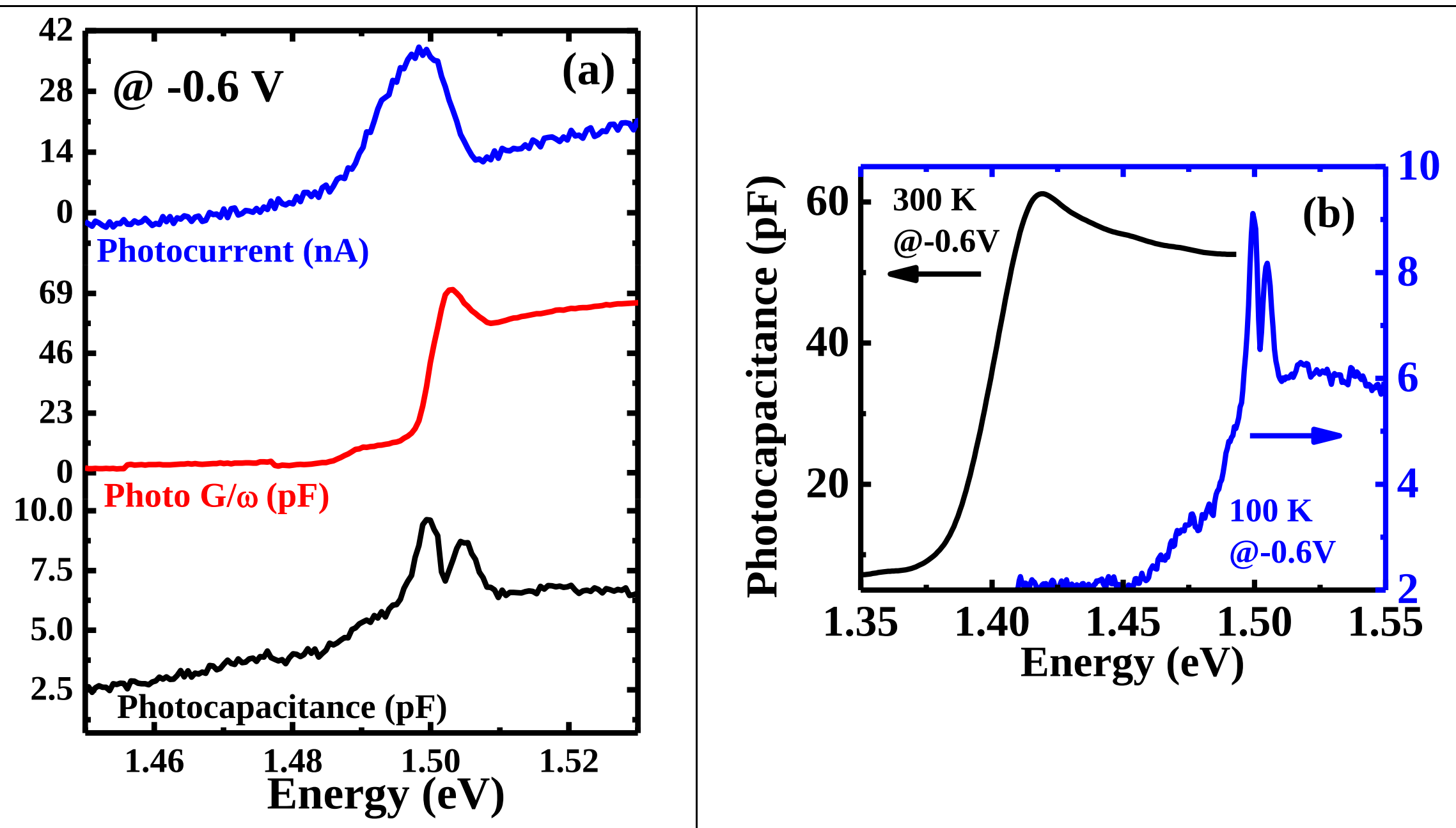

FIG. 4. (a) Photocapacitance and Photo-G/ω measured at 200 Hz is compared with DC-photocurrent Spectra for -0.6 V bias at100K. (b) Photocapacitance spectra at 100 K and 300 K are compared at a similar reverse bias of -0.6 V and frequency of 200 Hz. Spectral peak energy and spectral shape changed significantly with temperature.

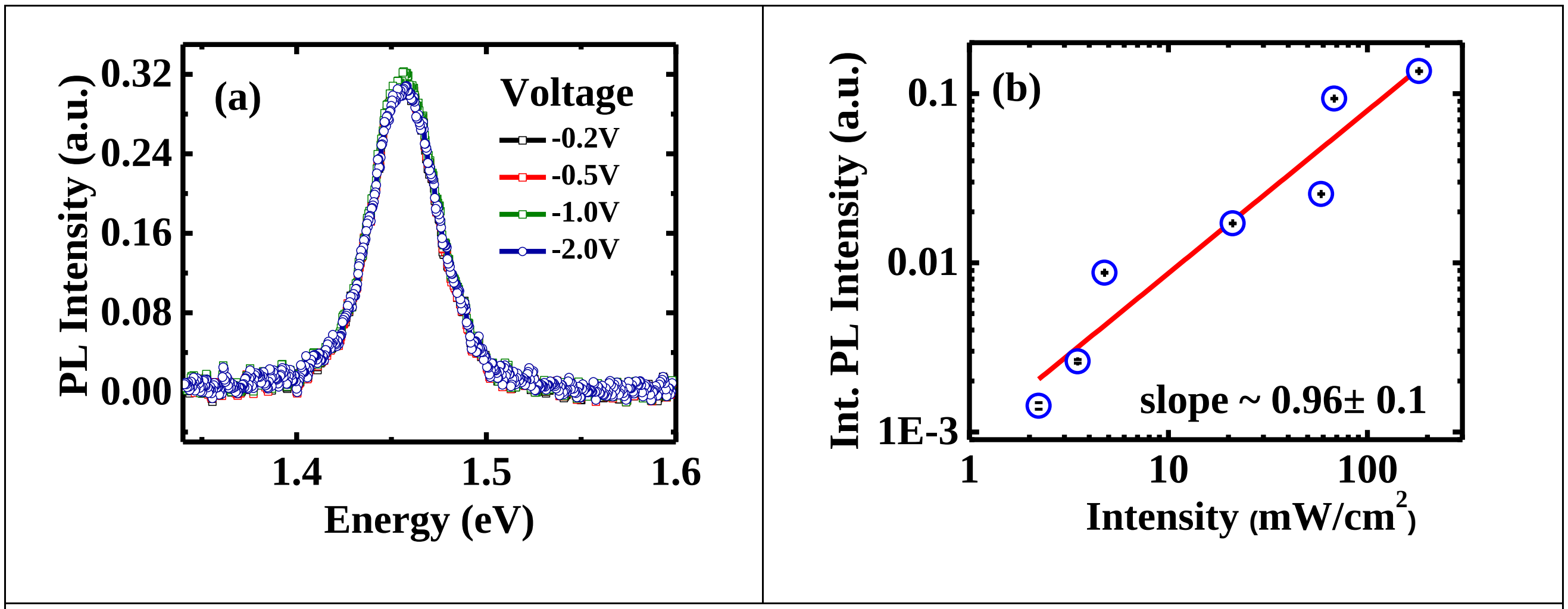

FIG. 5. (a) Photoluminescence Spectra at 100 K under different reverse biases. However, we did not observe any bias dependent spectral splitting and spectral red shift in the photoluminescence spectra. (b) Integrated PL intensity versus excitation intensity follows a linear variation in log-log plot which gives a slope ~0.96 at 100 K.